\documentclass[12pt]{article}
\usepackage{amssymb,amsmath,epsfig}
\usepackage[utf8]{inputenc}
 \allowdisplaybreaks
\begin{document}
\title{\bf Charged Compact Stars in Extended $f(\mathcal{R},\mathcal{G},\mathcal{T})$ Gravity}

\author{M. Ilyas$^1$ \thanks{ilyas\_mia@yahoo.com}, A. R. Athar$^2$ \thanks{athar\_chep@hotmail.com} and Asma Bibi$^1$ \thanks{asmawahid548@gmail.com}\\
$^1$ Institute of Physics, Gomal University,\\
Dera Ismail Khan, 29220, KP, Pakistan\\
$^2$ Institute of Physics, The Islamia University of Bahawalpur,\\
Baghdad-ul-Jadeed Campus, Bahawalpur-63100, Pakistan}

\date{}

\maketitle
\begin{abstract}
The purpose of this paper is to study charged compact stars using extended gravitational theory, also known as $f(\mathcal{R}, \mathcal{G}, \mathcal{T})$ gravity. Alternatively, this theory is also called $f(\mathcal{R}, \mathcal{T}, \mathcal{G})$ gravity. The symbols $\mathcal{R}, \mathcal{G}$, and $\mathcal{T}$ denote the Ricci Scalar, the Gauss-Bonnet invariant, and the trace of the energy-momentum tensor, respectively. We suggested several plausible models in the framework of this new gravity theory, and then used these models to explore several physical properties of compact objects of relativistic nature. This research also takes into account three famous compact stars: Vela X-1 (CS1); SAXJ1808.4-3658 (CS2); and 4U1820-30 (CS3). Moreover, using the suggested models, the physical nature of anisotropic stress, energy density, various energy conditions (ECs), the state of equilibrium, interior stability, mass variations, compactness, anisotropy, electric charge, and electric field intensity are analysed for considered compact stars. Different plots of the above-mentioned quantities are presented for this analysis. Conclusively, the ECs are satisfied, and the compact stars have a significant dense core.
\end{abstract}
{\bf Keywords:} $f(\mathcal{R,T,G})$ gravity; Compact stars; Stability.\\
{\bf PACS:} .

\section{Introduction}
Researchers have proposed various useful improvements to the General Relativity (GR) in the preceding years, despite its status as a great, well-established, and highly effective theory. Modified gravity theories (MGTs) have been demonstrated to be viable options for explaining phenomena like the existence of dark matter (DM) and the universe's accelerated expansion \cite{ref1BCS,ref2BCS,ref3BCSref15ref7.3CSI}. These theories discuss the problems that appear to be unsolved by the GR. From the other perspective, they require us to incorporate some as-yet unidentified fluid constituents like Dark Energy (DE) and DM and into the cosmos. Another of the sources of motivation for MGTs has been to figure out some of the challenges related to the infrared and ultra-violet scales that describe the distinct phases of the universe's accelerated expansion. The primary objective of MGTs is to expand GR by introducing $\mathcal{R}$ to the action. Since the inception of GR's theory of gravity, the notion of MGTs has been around owing to the anomalous instability \cite{ref4BCS}. Furthermore, it is a well-established fact that GR by itself does not convey the idea of the universe's expansion without the inclusion of some other component in the Lagrangian \cite{ref8BCSref12}. Since the emergence of MGTs, the high-energy physics, astrophysics, and cosmology communities have been attempting to find more information about it. In such modifications, instead of the $\mathcal{R}$, an arbitrary function is used. Examples of such functions are $f(\mathcal{R})$, $f(\mathcal{G})$, where $\mathcal{G}$ is the Gauss Bonnet (GB) invariant, and plenty others, as discussed in Refs: \cite{ref1,ref2,ref3,ref4,ref5,ref6,ref7,ref8,ref9}.\\
These modified theories are addressing the extraordinary phenomenon of the universe's expansion.\cite{ref10}. As a matter of fact, it is well documented that GR alone in its definite pattern cannot explain the acceleration of the cosmos without introducing an additional term in the gravitational Lagrangian ($L_{G}$) or exotic matter \cite{ref8BCSref12,ref11}. A basis modification that Buchdal made in 1970 \cite{ref13} was to change $\mathcal{R}$ in the Hilbert Einstein (HE) gravitational action to $f(\mathcal{R})$. The literature on MGTs is extensive \cite{ref3BCSref15ref7.3CSI,ref14,ref16,ref17,ref18,ref19,ref20,mia5,mia6}. One of the best explanations for the universe's accelerated expansion is $f(\mathcal{R})$ gravity \cite{ref3CSI}, which was originally suggested in \cite{ref4CSI} and also provided the prerequisites for developing a workable cosmological model \cite{ref5.1CSI,ref5.2CSI,ref5.3CSI,ref5.4CSIref34,ref5.5CSI,ref5.6CSI,ref5.7CSIref6CSI,ref5.8CSI} that is completely compatible with Newton's law.\\
In extended $f(\mathcal{R})$ gravity, a simple and general feasible model with a non-minimal coupling term between the matter and geometry parts was developed \cite{ref5.7CSIref6CSI}. The continuous accelerated expansion of the cosmos is described by this coupling factor, which is regarded as the basis of gravity. Some might explore additional information in Ref:\cite{ref3BCSref15ref7.3CSI,ref7.1CSI,ref7.2CSI}. Examining the diverse types of matter in action (Lagrangian density), this coupling generates an additional force \cite{ref8CSI}, so it can be clearly understood that a more appropriate and instinctive form of action need not necessitate eliminating this additional force \cite{ref9.1CSI,ref9.2CSI}. Additionally, the consequences of a non-minimal coupling on the relativistic stellar equilibrium situation were investigated.\\
Gauss Bonnet gravity (GBG), a well-known modified theory, has been explored extensively in recent years \cite{ref22ref14.1CSI,ref23}. In this extended gravitational theory's HE-action, a function $f(\mathcal{R}, \mathcal{G})$ is used. Besides, the invariant term of GB is referred to as:
\begin{equation*}
\mathcal{G}=\mathcal{R}_{\alpha\beta\xi\eta}\mathcal{R}^{\alpha\beta\xi\eta}-4\mathcal{R}_{\beta\xi}\mathcal{R}^{\beta\xi}+\mathcal{R}^2,
\end{equation*}
where in the above expression, $\mathcal{R}_{\alpha\beta\xi\eta}$ and $\mathcal{R}_{\beta\xi}$ represent Riemann tensor and Ricci tensor respectively. GBG is independent of spin-two ghost instabilities \cite{ref10.1CSI,ref10.2CSI,ref10.3CSI} and becomes the Lovelock scalar invariant of the second order.\\
When the GB component (4D topological invariant) is linked to scalar fields or expressed in terms of a $f(\mathcal{G})$ function, some interesting results can be found, while the field equations (FEqs)are not affected by this \cite{ref11.1CSI,ref11.2CSI,ref11.3CSI}. Nojiri and Odintsov developed a model based on arbitrary functions, which is regarded as $f(\mathcal{G})$-MGT \cite{ref21ref12CSI}. In the same way as previous MGTs, the said theory is a viable approach for studying DE that also integrates the limitations of the solar system \cite{ref13CSI}. This concept was used to analyse phase transitions, such as the transition from deceleration phase to acceleration phase and from non-phantom phase to phantom phase, along with the early and late-time unified expansion of the universe \cite{ref22ref14.1CSI,ref14.2CSI}. With the advent of MGTs, the striking issue of the universe's accelerated expansion is effectively explored.\\
Research into compact objects like BHs, neutron stars, and pulsars has lately been supported by the use of various MGTs to examine different aspects of these strange stars \cite{ref37ref15CSI,ref39ref16CSI,ref38ref17CSI,ref18CSI}. Spherical symmetric geometry, which allows for a wide range of matter distribution options, is a useful tool for developing plausible mathematical models. A perfect fluid distribution is often assumed in the investigation of compact objects, but anisotropic fluids are also often taken into account. Because of the anisotropy, the stability of the fluid is somewhat less stable than in a perfect fluid case. The implications of local anisotropy are investigated with the use of the equation of state (EoS) \cite{ref40ref19CSI}. Because of this, it becomes more reasonable to presume that the matter in MGTs models has an anisotropic composition. Mostly in the existence of anisotropic fluid and charge, several physical attributes of numerous compact stars have already been described \cite{ref41ref20CSI,ref42ref21CSI,ref43ref22CSI,ref44ref23CSI,mia1,mia2,mia3,mia4}.\\
As one example, consider the fact that, in the context of the enormous expansion of the universe, the additional GB term is able to correct some deficiencies of the $f(\mathcal{R})$ gravitational theory \cite{ref21ref12CSI,ref22ref14.1CSI,ref23,ref24,ref25}. In terms of $f(\mathcal{R}, \mathcal{G})$-MGT, $f(\mathcal{G})$ gravity is the most basic form that has been widely addressed and is capable of reproducing any type of cosmological solution.\\
For example, it might aid in the potential research of acceleration domains and subsequent transition to decelerated domains during the inflationary period, and it satisfies all conditions prompted by solar system investigations and bridges the phantom dividing line \cite{ref26,ref27}. It has been argued previously in Ref. \cite{ref28} that $f(\mathcal{G})$-gravity is less limited than $f(\mathcal{R})$-gravity. As an alternative to DE, the $f(\mathcal{G})$-gravity provides a convenient framework for the investigation of a variety of cosmic concerns \cite{ref29}. Similarly, in an accelerating universe, $f(\mathcal{G})$ gravitation is extremely useful in understanding the behaviour of time-constrained future singularities and late time epochs \cite{ref30,ref31}. As a result of various plausible models in $f(\mathcal{G})$-gravity, it is also examined that the cosmic accelerating essence is preceded by the matter epoch \cite{ref28,ref29}. For the goal of satisfying various specific solar system requirements, several feasible $f(\mathcal{G})$-gravity models have been proposed \cite{ref28,ref29}. Models like these are examined even more in \cite{ref32}, and new restrictions on $f(\mathcal{G})$-gravity models may come from how energy conditions (ECs) change \cite{ref33,ref5.4CSIref34,ref35}.\\
As a result of observations of compact phenomena such as BHs, pulsars, and neutron stars, scientists are now focusing their efforts on developing effective physical models based on very accurate observational data \cite{ref36}.\\
The compact star has received substantial attention from researchers in the last several decades. There is much to be studied and investigated regarding compact objects, but they are commonly considered to have large masses and small radii. This is what gives them the characteristic of being extremely dense entities. As it turns out, gravitational physics has made great progress in finding accurate solutions in the case of astronomical objects. A recent study looked into most of the physical characteristics of various strange compact stars in the context of multiple modified theories of gravity, and the results suggested that all of these objects are stable, have realistic matter content, and fulfil all of the ECs \cite{ref37ref15CSI,ref39ref16CSI,ref38ref17CSI}. The spherically symmetric geometry is especially effective and logical for representing static objects, but there are many more possibilities when it comes to matter content. In the past, a lot of studies concentrated on getting the fluid matter content to be ideal. Similarly, fluids with different viscosities and anisotropic pressure have been examined, with the conclusion that the anisotropy affects the stable configuration when compared to the local isotropic situation. In addition, the EoS has also been used to expound on the implications of local anisotropy \cite{ref40ref19CSI}. Because of this, models of modified gravity appear appropriate for dealing with anisotropic pressure. It is known that pressure anisotropy and charge can affect the physical parameters of compact stars \cite{ref41ref20CSI,ref42ref21CSI,ref43ref22CSI,ref44ref23CSI}.\\
In particular, in this study, we present a novel type of MGT known as $f(\mathcal{R}, \mathcal{G}, \mathcal{T})$-gravity, in which we developed the $L_{G}$ by substituting a generic function $f(\mathcal{R}, \mathcal{G}, \mathcal{T})$ for $\mathcal{R}$ in the HE-action. In Section 2 of this study, which includes the following structure, we construct the FEq and consider the anisotropic matter distribution with static spherical geometry. Next, we examine certain stable geometries in the interest of making our theory quite relevant. We take various compact stars as stable geometries and study properties such as stability, fluctuations in energy density, stresses, ECs, and many more. In the final section, we summarise and illustrate our conclusions.

\section{$f(\mathcal{R,G,T})$ Modified Gravitational Theory}
We consider the appropriate representation of action to be used in developing the $f(\mathcal{R,G,T})$-MGT \cite{ilyas2021compact} and obtaining its corresponding FEqs:
\begin{equation}\label{1}
\mathcal{S}_{f(\mathcal{R,G,T})}=\frac{\kappa^{-2}}{2}\int d^{4}x\sqrt{-g}[f(\mathcal{R,G,T})+\mathcal{L}_{m}+\mathcal{L}_{e}].
\end{equation}
In this equation, $\kappa$ symbolises the coupling constant, and the symbol $g$ represents the determinant of the metric tensor $(g_{\alpha\beta})$. Mathematically, the tensor of energy and momentum ($\mathcal{T}_{\gamma\delta}$) is represented as follows \cite{landau1971classical}
\begin{equation}\label{2}
\mathcal{T}_{\gamma\delta}=-\frac{2}{\sqrt{-g}}\frac{\delta(\mathcal{L}_{m}\sqrt{-g})}{\delta
g^{\gamma\delta}},
\end{equation}
which, in consequence, is equivalent to,
\begin{equation}\label{3}
\mathcal{T}_{\gamma\delta}=g_{\gamma\delta
}\mathcal{L}_{m}-2\frac{\partial\mathcal{L}_{m}}{\partial
g^{\gamma\delta }}.
\end{equation}
The desired FEqs of $f(\mathcal{R,G,T})$-MGT are given by the action variation expressed in Eq. (\ref{1}), as
\begin{align}\nonumber
&{\mathcal{G}_{\rho \sigma }} = \\\nonumber\label{5}
&\frac{1}{{{f_{\cal R}}\left( {{\cal R},{\cal G},{\cal T}} \right)}}\left[ {{\kappa ^2}({{\cal T}_{\rho \sigma }}+{ E}_{\rho \sigma })} \right. - \left( {{{\cal T}_{\rho \sigma }} + {\Theta _{\rho \sigma }}} \right){f_T}\left( {{\cal R},{\cal G},{\cal T}} \right) \\\nonumber
& + \frac{1}{2}{g_{\rho \sigma }}(f\left( {{\cal R},{\cal G},{\cal T}} \right) + \mathcal{R}{f_{\cal R}}\left( {{\cal R},{\cal G},{\cal T}} \right)) + {\nabla _\rho }{\nabla _\sigma }{f_{\cal R}}\left( {{\cal R},{\cal G},{\cal T}} \right)\\\nonumber
& - {g_{\rho \sigma }}\Box{f_{\cal R}}\left( {{\cal R},{\cal G},{\cal T}} \right) - (2\mathcal{R}{\mathcal{R}_{\rho \sigma }} - 4\mathcal{R}_\rho ^\xi {\mathcal{R}_{\xi \sigma }} - 4{\mathcal{R}_{\rho \xi \sigma \eta }}{\mathcal{R}^{\xi \eta }}\\\nonumber
& + 2\mathcal{R}_\rho ^{\xi \eta \delta }{R_{\sigma \xi \eta \delta }}){f_{\cal G}}\left( {{\cal R},{\cal G},{\cal T}} \right) - (2\mathcal{R}{g_{\rho \sigma }}{\nabla ^2} - 2\mathcal{R}{\nabla _\rho }{\nabla _\sigma } - 4{g_{\rho \sigma }}{\mathcal{R}^{\xi \eta }}{\nabla _\xi }{\nabla _\eta }\\
& - 4{\mathcal{R}_{\rho \sigma }}{\nabla ^2} + 4\mathcal{R}_\rho ^\xi {\nabla _\sigma }{\nabla _\xi } + 4\mathcal{R}_\sigma ^\xi {\nabla _\rho }{\nabla _\xi } + 4{\mathcal{R}_{\rho \xi \sigma \eta }}{\nabla ^\xi }{\nabla ^\eta })\left. {{f_{\cal G}}\left( {{\cal R},{\cal G},{\cal T}} \right)} \right],
\end{align}
$\mathcal{G}_{\rho\sigma}=\mathcal{R}_{\rho\sigma}-\frac{1}{2}g_{\rho\sigma}\mathcal{R}$ and ${ E}_{\rho \sigma }$ are the Einstein and Maxwell tensors, respectively.
\begin{equation}
 {E_{\mu \nu }} = \frac{{{g_{\mu \mu }}}}{2}\left[ { - {F^{\mu \alpha }}{F_{\alpha \nu }} + \frac{1}{4}\delta _\nu ^\mu {F^{\alpha \beta }}{F_{\alpha \beta }}} \right].
 \end{equation}
The FEqs of the GR can also be found conveniently by choosing $f(\mathcal{R,G,T})=\mathcal{R}$. The trace of the equation (\ref{5}) is given as,
\begin{align}\nonumber
&{\kappa ^2}({\cal T }+ E) - ({\cal T} + \Theta ){f_{\cal T}}({\cal R},{\cal G},{\cal T})  - {f_{\cal R}}({\cal R},{\cal G},{\cal T})R - 3\Box {f_{\cal R}}({\cal R},{\cal G},{\cal T})+ 2{\cal G}{f_{\cal G}}({\cal R},{\cal G},{\cal T})\\
&- 2\mathcal{R}{\nabla ^2}{f_{\cal G}}({\cal R},{\cal G},{\cal T}) + 4{R^{\rho \sigma }}{\nabla _\rho }{\nabla _\sigma }{f_{\cal G}}({\cal R},{\cal G},{\cal T})+ 2f({\cal R},{\cal G},{\cal T}) = 0,
\end{align}
while $\Theta=\Theta^{\rho}_{\rho}$. Furthermore, Eq. (\ref{5}) has a non-zero divergence, since
\begin{align}\label{5a}\nonumber
{\nabla ^\rho }{{\cal T}_{\rho \sigma }} &= \frac{{{f_T}({\cal R},{\cal G},{\cal T})}}{{{\kappa ^2} - {f_{\cal T}}({\cal R},{\cal G},{\cal T})}}\left[ {\left\{ {{{\cal T}_{\rho \sigma }+E_{\rho \sigma }} + {\Theta _{\rho \sigma }}} \right\}{\nabla ^\rho }\left\{ {\ln {f_{\cal T}}({\cal R},{\cal G},{\cal T})} \right\}} \right.\\
& - \left. {\frac{1}{2}{g_{\rho \sigma }}{\nabla ^\rho }{\cal T} + {\nabla ^\rho }{\Theta _{\rho \sigma }}} \right].
\end{align}
Eq.(\ref{3}) must be differentiated in order to devise a useful result for $\Theta_{\sigma\rho}$:
\begin{equation}\label{6}
\frac{\delta \mathcal{T}_{\sigma\rho}}{\delta g^{\xi\eta}}=\frac{\delta
g_{\sigma\rho}}{\delta g^{\xi\eta}}\mathcal{L}_{m}+g_{\sigma\rho}
\frac{\partial\mathcal{L}_{m}}{\partial
g^{\xi\eta}}-2\frac{\partial^2\mathcal{L}_{m}}{\partial
g^{\xi\eta}\partial g^{\sigma\rho}}.
\end{equation}
Making use of relations
\begin{equation}\nonumber
\frac{\delta g_{\sigma\rho}}{\delta
g^{\xi\eta}}=-g_{\sigma\mu}g_{\rho\nu}\delta_{\xi\eta}^{\mu\nu},\quad
\delta_{\xi\eta}^{\mu\nu}=\frac{\delta g^{\mu\nu}}{\delta
g^{\xi\eta}},
\end{equation}
where $\delta_{\xi\eta}^{\mu\nu}$ denotes the Kronecker symbol in its most generic form. Now, substituting Eq.(\ref{6}) which yields the following:
\begin{equation}\label{7}
\Theta_{\sigma\rho}=-2\mathcal{T}_{\sigma\rho}+g_{\sigma\rho}\mathcal{L}_{m}-2g^{\xi\eta}
\frac{\partial^{2}\mathcal{L}_{m}}{\partial g^{\sigma\rho}\partial
g^{\xi\eta}}.
\end{equation}
This is useful for evaluating the tensor $\Theta_{\sigma\rho}$.

\section{Spherical Anisotropic Fluids and Boundary Conditions}
It is our main objective to investigate how anisotropic stress influences the mathematical modelling of relativistic compact star cores. Taking this into account, we may consider the anisotropic fluid to be as follows:
\begin{align}\label{2l}
{\mathcal{T}_{\sigma \lambda }} = (\rho  + {P_t}){V_\sigma
}{V_\lambda } - {P_t}{g_{\sigma \lambda }} + \Pi{X_\sigma }{X_\lambda }.
\end{align}
In the above Eq. (\ref{2l}),  $\Pi=(P_r-P_t)$ while $P_r$ and $P_t$ denote radial and tangential pressures. Moreover, the four vectors of fluid are represented by ${V_\sigma}$ and $X_\sigma$. Assuming that the considered system lies in a frame that is not tilted, this implies that these quantities should fulfil the ${V^\sigma }{V_\sigma }=1$ and ${X^\sigma }{X_\sigma }=-1$ relationships. As a result, by taking $\mathcal{L}_m=\rho$, the equation Eq. (\ref{2}) may be incorporated as,
$${\Theta _{\sigma \lambda }} =  - 2{\mathcal{T}_{\sigma \lambda }} + \rho {g_{\sigma \lambda }}.$$
Using this approach, it is possible to represent the equation of motion (EoM) (\ref{5}) in the following way:
\begin{equation}\label{fieldequation}
{\mathcal{R}_{\sigma \lambda }} - \frac{1}{2}\mathcal{R}{g_{\sigma \lambda }} = \mathcal{T}_{\sigma \lambda
}^{\textit{eff}}.
\end{equation}
In this particular scenario, $\mathcal{T}_{\sigma \lambda}^{\textit{eff}}$ represents the effective energy momentum tensor, which is expressed as,
\begin{align}\nonumber
&\mathcal{T}_{\sigma \lambda }^{\textit{eff}} =\frac{1}{{{f_{\cal R}}\left( {{\cal R},{\cal G},{\cal T}} \right)}}\left[ {{\kappa ^2}({{\cal T}_{\sigma \lambda }}+E_{\sigma \lambda })} \right. - \left( {{{\cal T}_{\sigma \lambda }} + {\rho g _{\sigma \lambda }}} \right){f_T}\left( {{\cal R},{\cal G},{\cal T}} \right) \\\nonumber
& + \frac{1}{2}{g_{\sigma \lambda }}[f\left( {{\cal R},{\cal G},{\cal T}} \right) + \mathcal{R}{f_{\cal R}}\left( {{\cal R},{\cal G},{\cal T}} \right)] + {\nabla _\sigma }{\nabla _\lambda }{f_{\cal R}}\left( {{\cal R},{\cal G},{\cal T}} \right)\\\nonumber
& - {g_{\sigma \lambda }}\Box{f_{\cal R}}\left( {{\cal R},{\cal G},{\cal T}} \right) - (2\mathcal{R}{\mathcal{R}_{\sigma \lambda }} - 4\mathcal{R}_\sigma ^\xi {\mathcal{R}_{\xi \lambda }} - 4{\mathcal{R}_{\sigma \xi \lambda \eta }}{\mathcal{R}^{\xi \eta }}\\\nonumber
& + 2\mathcal{R}_\sigma ^{\xi \eta \delta }{R_{\lambda \xi \eta \delta }}){f_{\cal G}}\left( {{\cal R},{\cal G},{\cal T}} \right) - (2\mathcal{R}{g_{\sigma \lambda }}{\nabla ^2} - 2\mathcal{R}{\nabla _\sigma }{\nabla _\lambda } - 4{g_{\sigma \lambda }}{\mathcal{R}^{\xi \eta }}{\nabla _\xi }{\nabla _\eta }\\
& - 4{\mathcal{R}_{\sigma \lambda }}{\nabla ^2} + 4\mathcal{R}_\sigma ^\xi {\nabla _\lambda }{\nabla _\xi } + 4\mathcal{R}_\lambda ^\xi {\nabla _\sigma }{\nabla _\xi } + 4{\mathcal{R}_{\sigma \xi \lambda \eta }}{\nabla ^\xi }{\nabla ^\eta })\left. {{f_{\cal G}}\left( {{\cal R},{\cal G},{\cal T}} \right)} \right].
\end{align}
As an additional consideration, we examine the static spherical symmetric geometry using the line-element expressed as,
\begin{equation}\label{zz7}
d{s^2} = {e^{a_1}}d{t^2} - {e^{a_2}}d{r^2} - {r^2}\left[ {d{\theta ^2} + {{\sin }^2}\theta d{\phi ^2}} \right],
\end{equation}
where the functions: $a_1$ and $a_2$ have radial dependence. These functions are assumed mathematically as $a_1(r)=Br^2+C$ and $a_2(r)=Ar^2$ \cite{krori1975singularity}. The letters A, B, and C in this instance represent arbitrary constants. Various physical conditions are used to determine the values of these arbitrary constants. In our case, a relativistic geometry (Eq. \ref{zz7}) is postulated with an anisotropic distribution of matter. It is reasonable to rewrite $a_1$ and $a_2$ as more particular arrangements of their inputs. It is possible to figure out the measurements of these constant parameters by taking into account different measurements of the configurations of compact stellar.\\
In this scenario, the charge contribution is shown in $\rho$, $p_r$ and $p_t$. Furthermore, considering the EoS of quark matter,
\begin{equation}\label{forcharge}
{p_r} = \frac{1}{3}\left[ {\rho  - 4{B_g}} \right].
\end{equation}

\section{Matching condition and Different Models}
This section includes the study of a hyper-surface ($\Sigma$) which acts as a boundary for both (the interior and exterior) zones. This hypersurface's interior is characterised by the KroriBarua (KB) metric. It is important to note that KB-space-time lacks singularities. In addition, we assume the Reissner-Nordström metric to characterise exterior geometry, which is expressed as,
\begin{equation}
d{s^2} = \left[ {1 - \frac{{2m}}{r} + \frac{{{Q^2}}}{{{r^2}}}} \right]d{t^2} - {\left[ {1 - \frac{{2m}}{r} + \frac{{{Q^2}}}{{{r^2}}}} \right]^{ - 1}}d{r^2} - {r^2}\left[ {d{\theta ^2} + \sin {\theta ^2}d{\varphi ^2}} \right],
\end{equation}
here the three quantities, $Q$, $r$ and $m$ are referred to as charge, radius, and mass, accordingly. For the charged fluid distribution, the metric's (Eq. (\ref{zz7})) interior (KB) and the Reissner-Nordström metric's exterior are perfectly compatible. Now, mapping $r=R$ and $m(R)=M$ for both the aforementioned geometries, the following is obtained:
\begin{equation}
A = -\frac{{ 1}}{{{R^2}}}\ln \left[ {1+\frac{{{Q^2}}}{{{R^2}}} - \frac{{2M}}{R}} \right],
\end{equation}
\begin{equation}
B = \left( {\frac{M}{R^3} + \frac{{{Q^2}}}{{{R^4}}}} \right){\left[ {1  + \frac{{{Q^2}}}{{{R^2}}}- \frac{{2M}}{R}} \right]^{ - 1}},
\end{equation}
\begin{equation}
C = \ln \left[ {1  + \frac{{{Q^2}}}{{{R^2}}}- \frac{{2M}}{R}} \right] - \left( {\frac{M}{R} - \frac{{{Q^2}}}{{{R^2}}}} \right){\left[ {1  + \frac{{{Q^2}}}{{{R^2}}}- \frac{{2M}}{R}} \right]^{ - 1}}.
\end{equation}
\begin{table}[h!]
\begin{tabular}{|c| c| c| c| c| c| c|}
 \hline
Compact Stars &$M$ & $R(km)$ & $\mu_M=\frac{M}{R}$ &$A$ &$B$ &$\mu_C=\frac{Q^2}{R^2}$ \\ [0.5ex]
\hline\hline\
CS1  & $1.77M_{\odot}$ & $9.56$ & $0.273091$ & $0.00832706 $ &$0.00608302 $& $0.0133624$\\ [1ex]
\hline\
CS2& $1.435M_{\odot}$ & $7.07$ & $0.299$ & $0.0169456$ &$0.0127081$ & $0.0266898$\\ [1ex]
\hline
CS3& $2.25M_{\odot}$ & $10$ & $0.332$ & $0.00760739 $ &$0.00555676$ & $0.0133208$\\ [1ex]
\hline
\end{tabular}
\caption{The approximated masses, radii and compactness of CS1, CS2, and CS3 along with corresponding numerical constants $A$, and $B$.}
\label{table:1}
\end{table}
In Table. \ref{table:1}, we give the numerical values for these constants (A, B, and C) for three different strange stellar objects with compact physical structures based on their estimated masses ($M$) and radii ($R$).
Aside from that, we shall adopt some feasible models for the investigation of various compact star attributes, such as the ECs and stability analysis, among other factors. As,
$$f(\mathcal{R,G,T})=f_i(\mathcal{R,G})+\lambda \cal{T}.$$
In this case, we shall use three distinct model $i=1, 2, 3$.

\subsection{Different Models}
We must use plausible models in this theory for studying various cosmic aspects of $f(\mathcal{R,G,T})$-MGT and learn more about how compact stars are mathematically modelled. Many new cosmological discoveries at both astrophysical and theoretical scales could be made depending on how the $f(\mathcal{R,G,T})$ models used in current research are chosen.\\
Three alternative $f(\mathcal{R,G,T})$-MGT models will be used to fulfill this objective; for instance,
\begin{equation}
f(\mathcal{R,G,T})=f(\mathcal{R,G})+\lambda\mathcal{T}
\end{equation}
With this method that $f(\mathcal{R})=\mathcal{R} + \alpha \mathcal{R}^2$, it is possible to examine compact star viability for distinct values of $\alpha$ without using intermediary approximations in equations system, as shown in Ref. \cite{astashenok2017realistic}. The analysis shows that there is an exponential decrease in scalar curvature with respect to distance, for $\alpha>0$. But at the surface of the star, the value of scalar curvature does not go to zero, which is very close to what GR predicts. In consequence, when $\alpha$ rises, the stellar mass contained by star's surface reduces. According to Einstein's forecast, a reduction in the surface mass of compact stars could potentially result in a decreased surface red-shift in the $R^2$-gravity framework. Furthermore, by simply choosing $f(R)$-gravity, we can estimate the relevant matching conditions at the edge of compact stars without requiring the scalar curvature to disappear.
\begin{itemize}
  \item \textbf{Model-(I)}\\
In the $f(\mathcal{R,G,T})$-MGT backgrounds, assuming the model given below:
\begin{equation}\label{model1}
f(\mathcal{R,G,T})=\mathcal{R} + \alpha \mathcal{R}^2 + \beta \mathcal{G}^n + \gamma \mathcal{G} \ln(\mathcal{G})+\lambda \mathcal{T}.
\end{equation}
In the above model (\ref{model1}), the symbols $\alpha,\beta,\gamma,\lambda$ and $n$ denote model's parameters, which are constants.\\ The GB curvature scalar $\mathcal{G}^{n}$ is considered in this suggested model, where $n$ denotes a real constant incorporating a scale-invariant field equation. This is one of the possible parallels to the $f(\mathcal{R})$ theories. A constant $\beta$ multiplies with $\mathcal{G}^{n}$ in this proposed model. This model suggests multiplying $\mathcal{G}^{n}$ by a constant $\beta$. Furthermore, because of the constraint of a dimensionless logarithmic term, we should have chosen this term as $log(\frac{\mathcal{G}}{\mathcal{G}_{0}})$. In this case, by resetting the constant $\beta$, it is possible to avoid a change in $\mathcal{G}_{0}$ value, because the term including the constant $\beta$ value has no significance in the FEq. Classically, it is possible to enable the term that incorporates constant $\beta$ to be zero.\\

  \item \textbf{Model-(II)}\\
The second viable model in $f(\mathcal{R,G,T})$-MGT is taken as following:
\begin{equation}\label{model2}
f(\mathcal{R,G,T})= \mathcal{R} + \alpha \mathcal{R}^2 + \beta \mathcal{G}^{n_2} (1 + \gamma \mathcal{G}^{m_2})+\lambda \mathcal{T}.
\end{equation}
In this suggested model, $\alpha,\beta,\gamma,\lambda$, $n_2$ and $m_2$  represent model's constant parameters.\\
To avoid possible singularities, one needs to figure out the constraints on the model's parameters ($n_2,m_2, \beta$ and $\gamma$). It is challenging yet important to find out these conditions.
For $G\rightarrow \pm \infty$ or $G\rightarrow 0^{-}$, this $f(\mathcal{R,G,T})$-model may be expressed as $f(\mathcal{R,G,T})\sim \mathcal{R} + \alpha \mathcal{R}^2 +\beta G^{n_{3}}+\lambda \mathcal{T}$.
One should necessarily notice that in this case we are not considering the trivial case $n_2=m_2 $, and taking $n_2>0$. \\
  \item \textbf{Model-(III)}\\
In the same manner, we present another plausible model in $f(\mathcal{R,G,T})$-MGT:
\begin{equation}\label{model3}
f(\mathcal{R,G,T})= \mathcal{R} + \alpha \mathcal{R}^2 + (a_1 \mathcal{G}^{n_3} + b_1)/(a_2 \mathcal{G}^{n_3} + b_2)+\lambda \mathcal{T}.
\end{equation}
In this case, model's constant parameters are listed as: $\alpha,\beta,\lambda$, $a_1$, $a_2$, $b_1$, $b_2$ and $n_3$.
One cannot find singularities when $\mathcal{G}$ is very large, because they tend to be constant in this case (it has been proven in literature that ($R+constant$) does not include any singularity and hence consistent with the $\Lambda$-CDM model). However, the occurrence of singularities can not be ruled out completely in the case of $G\rightarrow 0^{-}$ for $n_3>1$.
\end{itemize}
It is critical to note that the last term ($\lambda\mathcal{T}$) in all three of the above models of $f(\mathcal{R,G,T})$-MGT is linear. Changing $\lambda$ in the last term may influence different physical attributes (i.e. mass, pressure, radius, and energy density) of compact celestial objects. On the other hand, in the case of a fixed central star energy density, the mass of neutron (and strange) stars can increase with $\lambda$. If the radius of a star is taken into account, for increasing $\lambda$, the radius of a neutron star increases while the radius of a strange star decreases. These suggested models have been chosen in a particular way to avoid complicated calculations. As a consequence, results of these models uncover various unknown cosmic phenomenon on both scales (theocratical and astrophysical) depending on the choices of aforementioned terms. Furthermore, the suggested models are adaptive for describing both the late and early universes, which is a significant advantage.\\
The condition is that $p_r(R)=0$. This gives us different values of the constant ($B_g$) in the cases of three distinct compact objects: CS1, CS2, and CS3. These values are presented in Table: \ref{table:2}.
\begin{table}[h!]
\centering
\begin{tabular}{|c| c| c| c|}
 \hline
Models  & $B_g$ for CS1 & $B_g$ for CS2 & $B_g$ for CS3 \\ [0.5ex]
\hline\hline\
Model-I & 0.0016647 & 0.00314065 & 0.00152135\\ [1ex]
\hline\
Model-II & 0.0016647 & 0.00314065 & 0.00152135\\ [1ex]
\hline
Model-III & 0.00166468 & 0.00314017 & 0.00152134\\ [1ex]
\hline
\end{tabular}
\caption{The estimated $B_g$ values for CS1, CS2, and CS3 using suggested models.}
\label{table:2}
\end{table}
\\
By making use of these suggested models with Eq. (\ref{fieldequation}), we acquire $\rho$, $p_r$ and $p_t$ from which we analyse the various attributes of compact stars as presented in Table \ref{table:1}. In the following section, we will go through each of these attributes in detail.

\section{Aspects of $f(\mathcal{R,G,T})$ Gravity Models}
The physical attributes of the aforementioned charged compact objects from the interior solution are described in this section. Three distinct $f(\mathcal{R,G,T})$ feasible models are used for investigating the stability and anisotropic behaviour of these charged compact objects. The next subsections contain a detailed discussion of these attributes one by one.
\subsection{Anisotropic Stresses and Energy Density Variations}
\begin{figure}[h!]
\centering
\epsfig{file=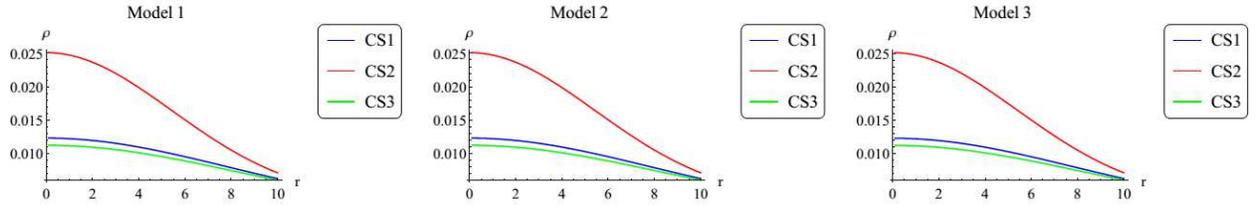,width=1\linewidth}
\caption{Variation of density profile for charged stars: CS1, CS2, and CS3, under different viable $f(\mathcal{R,G,T})$-models.}\label{roc}
\end{figure}

\begin{figure}[h!]
\centering
\epsfig{file=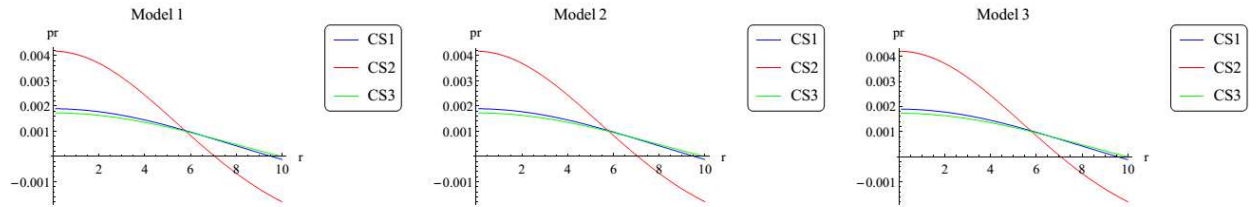,width=1\linewidth}
\caption{Evolution of radial pressure for charged stars: CS1, CS2, and CS3, under different plausible $f(\mathcal{R,G,T})$-models.}\label{prc}
\end{figure}

\begin{figure}[h!]
\centering
\epsfig{file=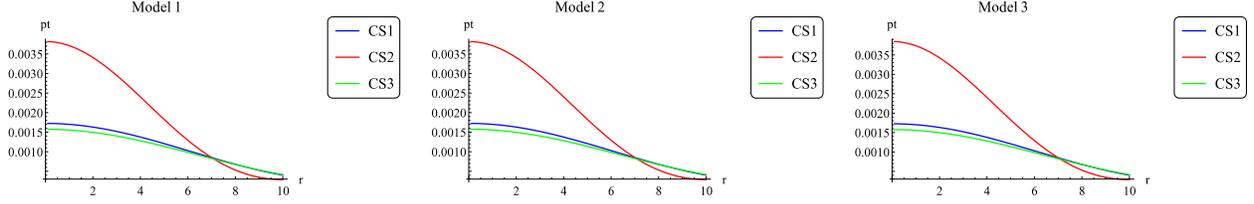,width=1\linewidth}
\caption{Transverse pressure's evolution for charged stars: CS1, CS2, and CS3, under different plausible $f(\mathcal{R,G,T})$-models.}\label{ptc}
\end{figure}

Using $f(\mathcal{R,G,T})$ MG-models, we investigate the impact of EoS for quark matter on anisotropic stresses at the core. Figures. (\ref{roc}, \ref{prc}) and (\ref{ptc}) demonstrate the relevant fluctuations in the proximity of energy density as well as anisotropic stresses.\\
Fig. (\ref{roc}) depicts the density evolution of three strange star candidates: CS1, CS2, and CS3. The density approaches its greatest value when $r$ approaches zero ($r\rightarrow0$). This also demonstrates the significant compactness of these stars' cores, confirming that our suggested $f(R,G,T)$-MGT models are compatible to the extent of the exterior region of the star's center.\\
The proposed Model-I suggests different values for density in the case of three distinct aforementioned strange star candidates. These values are given as $6.61955\times10^{14} gcm^{-3} $, $1.34564\times10^{15} gcm^{-3} $, and $6.04811\times10^{14} gcm^{-3}$ corresponding to CS1, CS2, and CS3 respectively.\\
In addition to above, considering Model-I, the suggested values of density at surface of the CS1, CS2, and CS3 are given as
$3.56864\times10^{14} gcm^{-3}$, $6.73266\times10^{14} gcm^{-3}$, and $3.26135\times10^{14} gcm^{-3}$ respectively.\\
We have presented corresponding values of density at the core and also at the surface using distinct models. Table \ref{table:3} shows these values.\\
Accordingly, Figs. (\ref{prc}-\ref{ptc}) demonstrate the fluctuation of radial and traverser pressures.\\
Furthermore, for $r\rightarrow0$, Model-I suggests different values of radial pressure at the center of three strange stars. The radial pressure at core of the CS1, CS2, and CS3 are $9.14009\times10^{34} g cm^{-1}sec^{-2}$, $2.01434\times10^{35} g cm^{-1}sec^{-2}$, and $8.34871\times10^{34} g cm^{-1}sec^{-2}$ respectively.\\
Similarly, Model-I suggests different values of transverse pressure, which are $8.30605\times10^{34} g cm^{-1}sec^{-2}$, $1.83924\times10^{35} g cm^{-1}sec^{-2}$, and $7.586\times10^{34} g cm^{-1}sec^{-2}$ corresponding to CS1, CS2, and CS3 respectively.
\\
The fluctuations of radial derivatives of density ($\frac{d\rho}{{dr}}$), radial pressure ($\frac{dp_r}{{dr}}$), and transverse pressure ($\frac{dp_t}{{dr}}$) can also be illustrated in the same manner. One can check that the aforementioned fluctuations are negative. Therefore, in the case of $r =0$, it implies
$${\left. {\frac{{d\rho }}{{dr}}} \right|_{r = 0}} = 0$$
$${\left. {\frac{{d{p_r}}}{{dr}}} \right|_{r = 0}} = 0$$
This is to be anticipated. The density at the core of a star, for example, is $\rho(r=0)=\rho_c$.
\begin{table}[h!]
\begin{tabular}{|c| c| c| c|}
 \hline
Compact Stars's Density & Model-I& Model-II & Model-III \\
\hline\hline\hline\
CS1 $\rho_c (g/cm^3)$ &$6.61955\times10^{14}$& $6.61955\times10^{14}$& $6.62184\times10^{14}$\\
\hline\
CS1 $\rho_R (g/cm^3)$ &$3.56864\times10^{14}$ &$3.56864\times10^{14}$& $3.56859\times10^{14}$\\
\hline\hline\
CS2 $\rho_c (g/cm^3)$ &$1.34564\times10^{15}$ &$1.34564\times10^{15}$& $1.34969\times10^{15}$\\
\hline\
CS2 $\rho_R (g/cm^3)$ &$6.73266\times10^{14}$ &$6.73266\times10^{14}$ &$6.73163\times10^{14}$\\
\hline\hline\
CS3 $\rho_c (g/cm^3)$ &$6.04811\times10^{14}$ &$6.04811\times10^{14}$& $6.0497\times10^{14}$\\
\hline\
CS3 $\rho_R (g/cm^3)$ &$3.26135\times10^{14}$& $3.26135\times10^{14}$& $3.26132\times10^{14}$\\
\hline
\end{tabular}
\caption{The Central ($\rho_c$) and Surface ($\rho_R$) densities approximations under suggested models. }
\label{table:3}
\end{table}
\subsection{Energy Conditions}
\begin{figure}[h!]
\centering
\epsfig{file=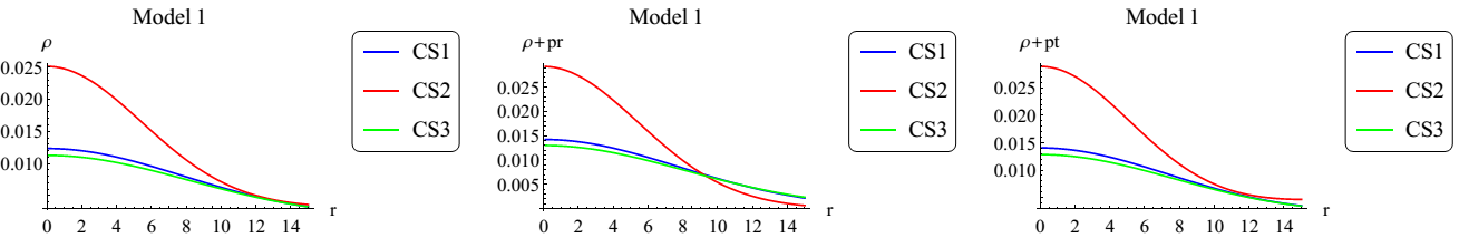,width=1\linewidth}
\epsfig{file=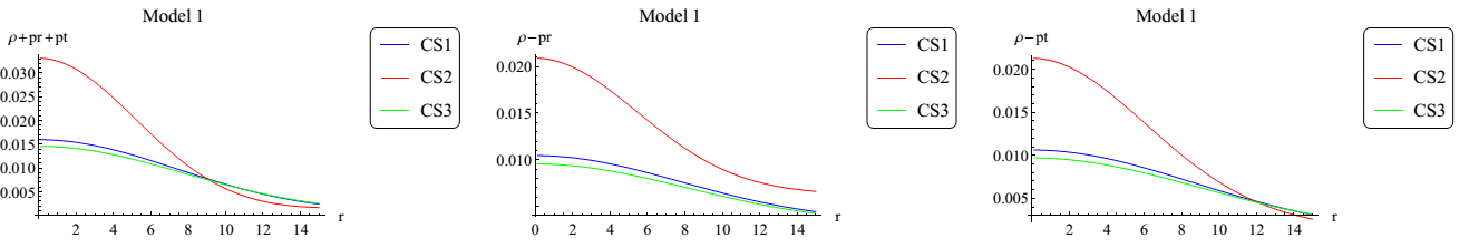,width=1\linewidth}
\caption{The various ECs for Model-I}\label{energy1}
\end{figure}

The ECs are mathematical limitations that must be fulfilled by the stress-energy tensor in order to effectively interpret a realistically feasible and admissible matter field. These coordinate invariant ECs are expressed as follows.\\
\begin{itemize}
\item NEC (\emph{Null Energy Condition}): $ \rho  + {p_i} \ge 0$ .\\
\item WEC (\emph{Weak Energy Condition}): $ \rho  \ge 0$, $ \rho  + {p_i} \ge 0$ .\\
\item SEC (\emph{Strong Energy Condition}): $ \rho  + {p_i} \ge 0$, $ \rho  + {p_i} + {p_t} \ge 0$ .\\
\item DEC (\emph{Dominant Energy Condition}): $ \rho  \ge |{p_i}|$ .
\end{itemize}
In the ECs listed above, $i=r, t$, and electric charge inputs are also incorporated into $\rho$, $p_r$ and $p_t$.\\
The investigation of suggested plausible $f(\mathcal{R,G,T})$ MG-models satisfies all of the aforementioned ECs for three distinct relativistic compact stars.
Fig. (\ref{energy1}) pictorially illustrate the transformation of these ECs in case of Model-I. Similar transformation of the ECs using Model-II and Model-II can be illustrated as well.
\subsection{The State of Equilibrium}
We employ the extended Tolman-Oppenheimer-Volko (TOV) equation to examine the equilibrium of the internal construction of these charged compact objects.
In case of charged and spherically anisotropic celestial interiors,the TOV equation is expressed as follows:
\begin{equation}\label{tov1}
\frac{{d{p_r}}}{{dr}} + \frac{{\nu'(\rho  + {p_r})}}{2} + \frac{{2({p_r} - {p_t})}}{r} + \frac{\lambda }{{3\left( {1 + 2\lambda } \right)}}\frac{d}{{dr}}\left( {3\rho  + {P_r} - 2{P_t}} \right)+ \frac{{\sigma Q}}{{{r^2}}}{e^{\lambda /2}} = 0
\end{equation}
The charge density is represented by $\sigma$. In addition, the aforementioned Eq. (\ref{tov1}) can possibly be written in terms of many forces, such as electric ($F_e$), gravitational ($F_g$), anisotropic ($F_a$), extra force ($F_{Ext}$) due to matter coupling curvature, and hydrostatic ($F_h$) forces.
\begin{equation}
F_g + F_h + F_a+F_{Ext}+F_e= 0,
\end{equation}
this results in
$$F_g=- r B(\rho+p_r),  F_h=\frac{{-d{p_r}}}{{dr}},   F_a= 2\frac{{({p_r} - {p_t})}}{r},$$ $$F_{Ext}= \frac{\lambda }{{3\left( {1 + 2\lambda } \right)}}\frac{d}{{dr}}\left( {3\rho  + {P_r} - 2{P_t}} \right), F_e= \frac{{\sigma Q}}{{{r^2}}}{e^{\lambda /2}}.$$
\begin{figure}[h!]
\centering
\epsfig{file=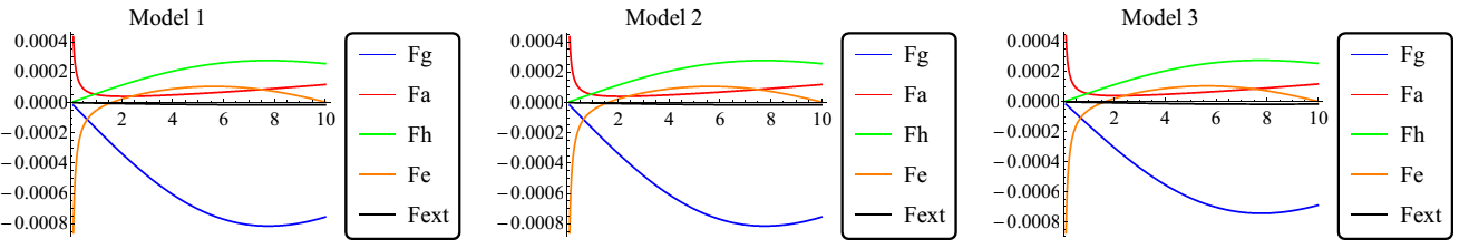,width=1\linewidth}
\caption{The variations of $F_h$, $F_g$, $F_a$, $F_{Ext}$ and $F_e$ for CS-1 in framework of suggested feasible $f(\mathcal{R,G,T})$-models.}\label{eqb}
\end{figure}
\begin{figure}[h!]
\centering
\epsfig{file=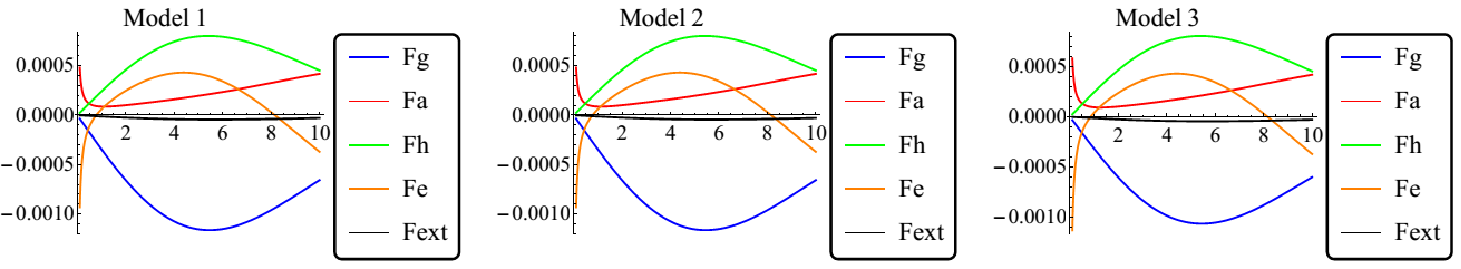,width=1\linewidth}
\caption{The variations of $F_h$, $F_g$, $F_a$, $F_{Ext}$ and $F_e$ for CS-2 in framework of suggested feasible $f(\mathcal{R,G,T})$-models.}\label{eqb}
\end{figure}
\begin{figure}[h!]
\centering
\epsfig{file=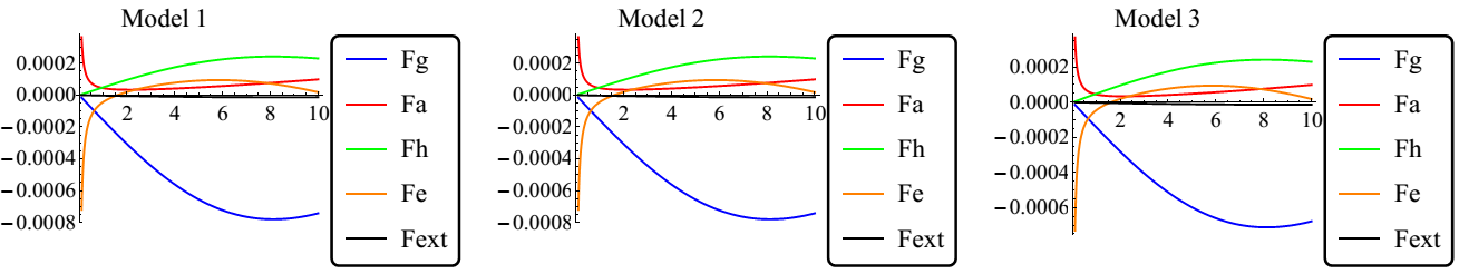,width=1\linewidth}
\caption{The variations of $F_h$, $F_g$, $F_a$ , $F_{Ext}$ and $F_e$ for CS-3 in framework of suggested feasible $f(\mathcal{R,G,T})$-models.}\label{eqb}
\end{figure}\\
We examine the hydrostatic equilibrium and variations of $F_g$, $F_h$, $F_a$, and $F_e$ forces, as illustrated in Fig. (\ref{eqb}), by applying the above descriptions to the values of various parameters from Table \ref{table:1}.\\
In Fig. (\ref{eqb}), variations of these forces are illustrated in the context of three suggested models. The left, centered, and right plots show these fluctuations in the background of models I, II, and III respectively. As seen in Fig. (\ref{eqb}), the $F_e$ has a relatively minor impact on this state of equilibrium.
\subsection{The Survey of Stability}
The analysis for star interiors stability using $f(\mathcal{R,G,T})$-MGT is presented in this section. It should be highlighted that only certain stellar models that are stable under fluctuations are important for the mathematical modelling of compact stellar formations. As a result, the importance of stability in the modelling of compact stars is an important research topic. Numerous studies have looked into the stabilisation of stellar structural systems. In present scenario, we use approaches that are premised on cracking (or overturning) ideas \cite{ref30}.
The radial ($v_{sr}^{2}$) and transverse ($v_{st}^{2}$) speeds of sound must be bracketed by a closed interval [0, 1] to satisfy the condition of causality, and the required condition ($0\le v_{sr}^{2}-v_{sr}^{2}\le1$) must be followed for stability.\\
The $v_{sr}^{2}$ and $v_{st}^{2}$ are mathematically expressed as:
$$\frac{{d{p_r}}}{{d\rho }} = v_{sr}^2$$
and
$$\frac{{d{p_t}}}{{d\rho }} = v_{st}^2.$$
The $v_{sr}^2\sim 1/3$ and $v_{st}^2$ satisfy the $0 \le v_{sr}^2 \le 1$ and $0 \le v_{st}^2 \le 1$ conditions. These exhibit that the casuality is preserved inside these charged compact objects.
\begin{figure}[h!]
\centering
\epsfig{file=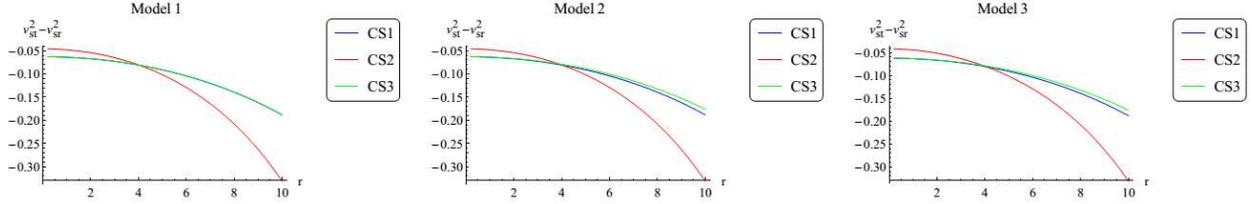,width=1\linewidth}
\caption{The fluctuations of $v_{st}^2 - v_{sr}^2$ in framework of suggested feasible $f(\mathcal{R,G,T})$-models.}\label{vstmvsr}
\end{figure}
In Fig. (\ref{vstmvsr}), we have presented plot of $v_{st}^2-v_{sr}^2$ for stability. It should be noted that under the exploration of several plausible $f(\mathcal{R,G,T})$-MGT models, all of our considered stellar configurations follow the constraint:
$$0< |v_{st}^2-v_{sr}^2|<1$$
In the current study, we found that $f(\mathcal{R,G,T})$-MGT models are stable for three examined strange candidate stars, which are: CS1, CS2, and CS3.
\subsection{Parameters of the State-Equation}
\begin{figure}[h!]
\centering
\epsfig{file=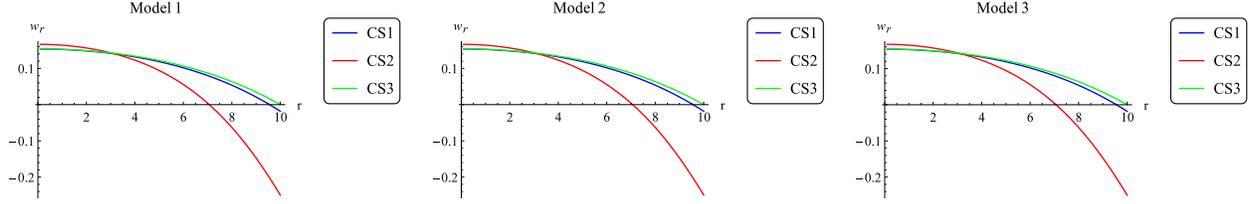,width=1\linewidth}
\caption{The behaviour of radial EoS-parameter under suggested feasible $f(\mathcal{R,G,T})$-models.}\label{wr}
\end{figure}

\begin{figure}[h!]
\centering
\epsfig{file=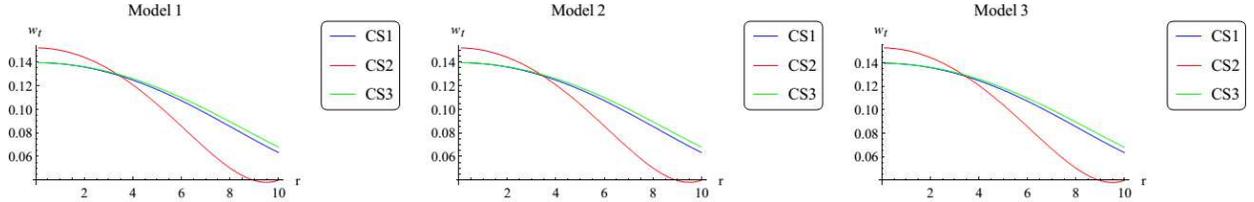,width=1\linewidth}
\caption{The behaviour of transverse EoS-parameter under suggested feasible $f(\mathcal{R,G,T})$-models.}\label{wt}
\end{figure}

There are two EoS parameters for anisotropic stresses, expressed as,
$$w_r= \frac{p_r}{\rho},$$
and
$$w_t =\frac{p_t}{\rho}.$$
The parameters of EoS must be between 0 and 1 in a radiation dominating epoch. Specifically, $0<w_r<1$ and $0<w_t<1$. In this study, we examine how EoS parameters evolve for three distinct compact stars, whose behaviour is pictorially depicted in Figs. (\ref{wr} and \ref{wt}).\\
It is clearly noticeable that $w_r$ and $w_t$ demonstrate the prescribed range.
\subsection{The Study of Mass-Radius and Compactness-Redshift Relations}
\begin{figure}[h!]
\centering
\epsfig{file=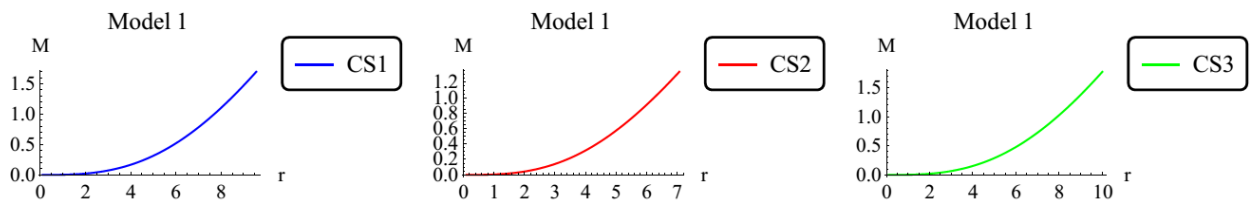,width=1\linewidth}
\caption{The $m(r)$- variations for CS1, CS2, and CS3 under Model 1.}\label{mass}
\end{figure}

\begin{figure}[h!]
\centering
\epsfig{file=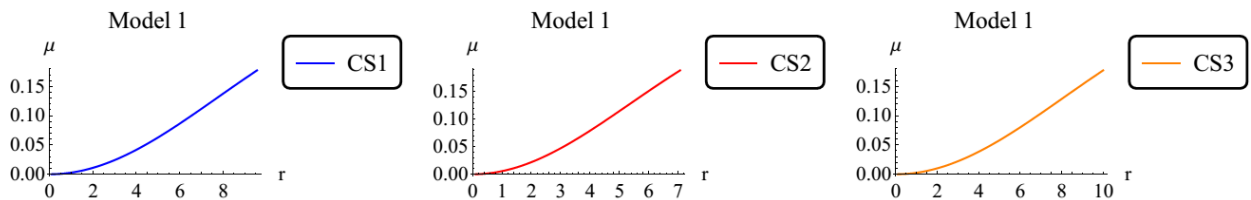,width=1\linewidth}
\caption{The $\mu(r)$- variations for CS1, CS2, and CS3 under Model 1.}\label{compact}
\end{figure}

The mass ($m$) of charged compact objects can be mathematically expressed in the following way:
\begin{equation}
m(r) = \int\limits_0^r {4\pi {{r'}^2}\rho dr'}.
\end{equation}
As the $m$ is a function of variable $r$, for $r=0$ the function $m(r)=0$, and for $r=R$ the function $m(r)=M$. The variations of the mass of charged compact stars is illustrated in Fig. (\ref{mass}).
We can observe that $m$ is consistent at the centre since it is in direct proportion with the radial distance, for example, when $r\rightarrow0$ the function $m(r)\rightarrow0$.
As depicted in Fig. (\ref{mass}), the mass function $m(r)$ reaches maximum for $r=R$. The mass radius relationship may also be used to investigate neutron stars in background of the $f(\mathcal{R,G,T})$-gravity.\\
Moreover, the compactness ($\mu$) is given as,
\begin{equation}
\mu (r) = \frac{1}{r}\int\limits_0^r {4\pi {{r'}^2}\rho dr'}
\end{equation}
Fig. (\ref{compact}) illustrates the $\mu$ variations for three distinct strange objects. Likewise, the Redshift ($Z_s$) of a compact star is mathematically defined as,
$$Z_s = {\left( {1 - 2\mu } \right)^{ - \frac{1}{2}}} - 1.$$
The $Z_s\leq2$ is constraint.
\begin{figure}[h!]
\centering
\epsfig{file=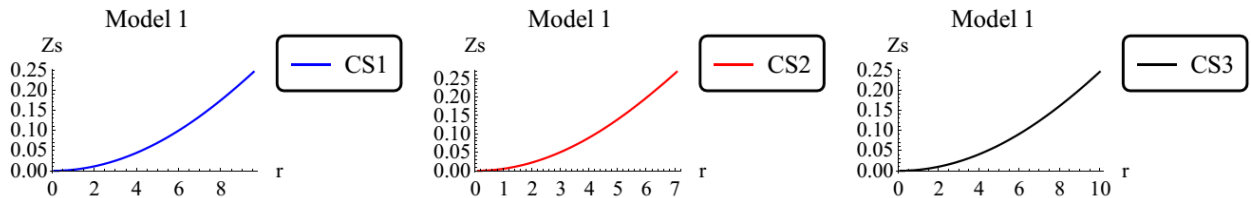,width=1\linewidth}
\caption{The $Z_s$-variations for CS1, CS2, and CS3 under Model 1.}\label{zs}
\end{figure}
In this scenario, we are studying at how $Z_s$ varies from the center to the surface of stars. These variations are shown using plots, as seen in Fig. (\ref{zs}).
\subsection{The Measurement of Anisotropy}
\begin{figure}[h!]
\centering
\epsfig{file=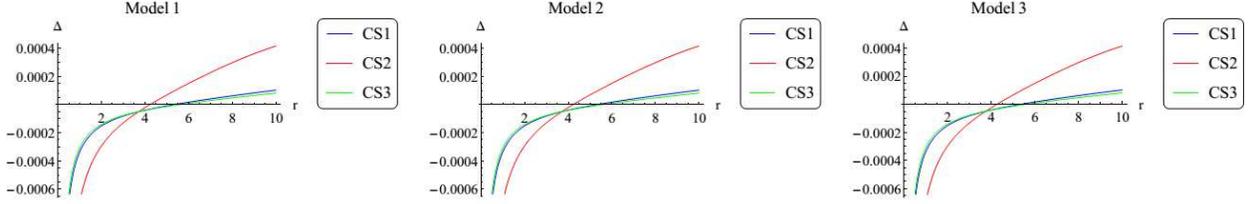,width=1\linewidth}
\caption{The $\Delta$-variations  w.r.t. radial coordinate ($r$).}\label{ani}
\end{figure}

It is critical to consider anisotropy in the modelling of relativistic star internal structures. Mathematically, anisotropy is expressed as:
\begin{equation}
\Delta  = \frac{2}{r}({p_t} - {p_r})
\end{equation}
In this study, we used suggested plausible models of $f(\mathcal{R,G,T})$-MGT to investigate $\Delta$ for three distinct charged strange objects.
We graph the anisotropy by inserting the constants into the suggested models and obtain that $\Delta > 0$ e.g. $p_t > p_r$ for $r$ becomes large.
This means that for all three considered compact objects, the anisotropy is oriented outward with respect to increasing $r$.
Fig. (\ref{ani}) depicts these plots.
It is worth noting that $\Delta\rightarrow 0$ at $r \rightarrow 0$ and with increasing $r$ around the star's surface, it becomes monotonously increasing outwards.
\subsection{The Electric Field and Charge}
\begin{figure}[h!]
\centering
\epsfig{file=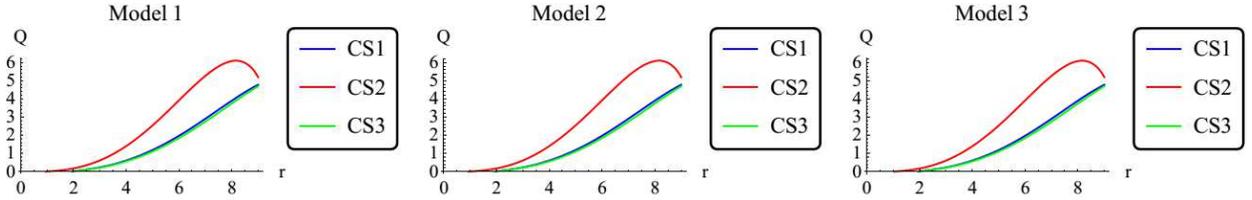,width=1\linewidth}
\caption{The $Q$-variations w.r.t. radial  coordinate.}\label{charg}
\end{figure}

\begin{figure}[h!]
\centering
\epsfig{file=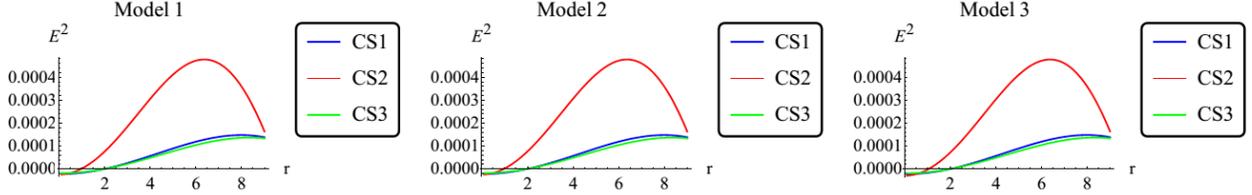,width=1\linewidth}
\caption{The $E^2$-variations w.r.t. radial coordinate.}\label{electric}
\end{figure}

Using suggested models, we computed values of the electric charge on the boundaries of three compact stars. In case of Model-I, values of electric charge are $7.91828\times10^{20}C$, $7.5532\times10^{20}C$, and $8.27752\times10^{20}C$ for Star 1,2, and 3 respectively while it is zero at the center of stars. Fig. (\ref{charg}) shows that at a distance from the center of these objects, values of the electric charge monotonously increase. The electric charge density of these stars decreases monotonously outward, maximizing at the core.
\\
Furthermore, we discussed the behaviour of electric field intensity ($E^2$), as illustrated in Fig. (\ref{electric}), which depicts fluctuations in $E^2$ for various compact objects.
\\
Keeping in view the plots, we deduce that at the center ($r=0$) of considered compact objects,$Q=0$, $\sigma=\sigma_0$ while at star's surface ($r=R$) $Q=Q_m$, $\sigma=0$ and $E^2=E^2_m$. It should be noted that in aforementioned expressions, $E^2_m$, $Q_m$, and $\sigma_0$ denote maximum electric field intensity, charge, and charge density respectively.\\
In comparison to GR-theory, the creation of stellar formations in the background of MGTs has a significantly larger contraction in the
collapse rate of spherical systems in its early phases. The presence of additional curvature (non-gravitational fluid) on the emergence
of compact formation may result in a more
compact realm comparing to GR-theory.\\
Analogously, the implications of these extra dark source factors on mass-radius relations for compact objects suggest extensive massive relativistic structures with radii that are relatively shorter when compared to GR-theory. Maybe the predicted observed masses of neutron-star models in MGTs shows that these stars are quite massive and have shorter radii when compared to GR-theroy. Such findings might provide a useful theoretical method for dealing with and studying huge and supermassive objects at enormous levels.

\section{Summary and Conclusions}
In the current research, we proposed a unique and more generic $f(\mathcal{R}, \mathcal{G},\mathcal{T})$-MGT. According to this theory, matter and geometry are coupled arbitrarily. The Lagrangian is constructed in this manner using a modified HE-action in which $\mathcal{R}$ is substituted by the arbitrary function $f(\mathcal{R}, \mathcal{G}, \mathcal{T})$. Next, using the variational principle technique, we produced the corresponding FEqs. The covariant divergence of $\mathcal{T}_{\gamma\delta}$ was found to be non-zero, agreeing with $f(\mathcal{R}, \mathcal{T})$-MGT \cite{harko2011f}. Furthermore, we analysed the influence of the suggested $f(\mathcal{R}, \mathcal{G},\mathcal{T})$-models by considering several compact configurations. To accomplish this goal, the physical properties of three compact objects (CS1, CS2, and CS3) are determined. The following are some of the most important findings:
According to our findings, energy density, $P_r$, $P_t$, and anisotropic stresses approach their maximum value at the core of considered compact stars, while they are at their minimum at the surface. It is important to note that the $P_r$ at the surface of the considered compact objects is zero. When dealing with the core of these stars, $rho_c$ approaches a significantly enormous value, making the stars very compact. The high compactness provides an acceptable justification for the validation of our suggested $f(\mathcal{R}, \mathcal{G},\mathcal{T})$-models.\\
In the current research, we also examined the influence of additional force (due to high curvature) and found a significantly realistic distribution of matter. Additionally, considering these compact stars, we determined that various ECs are fulfilled, and this also suggests the normal distribution of matter (no exotic matter). The importance and relevance of these ECs in the background of MGTs can be found in Refs. \cite{capozziello2015generalized,capozziello2014energy} and in the framework of non-local gravity in Ref. \cite{ilyas2019energy}.\\
It is noteworthy that for modelling any compact object, analysis of the stability factor is very important. In this regard, cracking methods are adopted to investigate the stability. The stability analysis established that our suggested models are conceivably reliable against the variations.\\
Different distinct forces ($F_a, F_g$, $F_{Ext}$, $F_e$ and $F_h $) emerge, balancing each other. The $F_g$ are directed outward while $F_a$, $F_{e}$, $F_{h}$ and $F_{Ext}$ (repulsive) is directed inward. The $F_g$ cancel out the combined effects of all these forces. The combination of these three forces accomplishes the equilibrium state. While studying the anisotropy, two separate EoS parameters ($w_r, w_t$) contribute. These both Eos parameters range in limit of realistic and normal distribution of matter.\\
As a result, we conclude that the $f(\mathcal{R}, \mathcal{G}, \mathcal{T})$-models we propose are valid and provide adequate justifications for physical properties. The entire investigation was conducted in conjunction with a proper contrast of numerous compact star candidates, confirming the viability of our suggested models. Moreover, our obtained results also agree with the results acquired using different MGTs [17,35,36,37]. Therefore, the implications of our adopted methods lead to a more appropriate explanation of the compact objects.

\section{Appendix A}
We suppose a specific arrangement of $f(\mathcal{R,G,T})$-MGT as $f(\mathcal{R,G,T})=f(\mathcal{R,G})+\lambda \mathcal{T}$. Then for metric Eq. (\ref{zz7}), we solve FEq (\ref{fieldequation}) to obtain following,
\begin{align}\nonumber
\rho & =  - \frac{1}{{8\left( {2\lambda  + {f_R} + 1} \right)}}((\frac{\lambda }{{\lambda  + {f_R} + 1}} + 1)(\frac{{{Q^2}}}{{\pi {r^4}}} - 8{{\rm{e}}^{ - b}}{r^{ - 2}}\left( {rb' + {{\rm{e}}^b} - 1} \right){f_R}^2\\\nonumber
& + 4f - 4R{f_R} - 4G{f_G}{f_R} - 4{{\rm{e}}^{ - b}}b'G'{f_{RG}} + {r^{ - 1}}16{{\rm{e}}^{ - b}}G'{f_{RG}} + 8{{\rm{e}}^{ - b}}{G^{\prime \prime }}{f_{RG}} + 8{{\rm{e}}^{ - b}}\\\nonumber
&{{G'}^2}{f_{RGG}} - 4{{\rm{e}}^{ - b}}b'R'{f_{RR}} + 16{r^{ - 1}}{{\rm{e}}^{ - b}}R'{f_{RR}} + 8{{\rm{e}}^{ - b}}{R^{\prime \prime }}{f_{RR}} + 16{{\rm{e}}^{ - b}}G'R'{f_{RRG}}\\\nonumber
& + 1/{r^2}16{{\rm{e}}^{ - 2b}}{f_R}(({{\rm{e}}^b} - 3)b'\left( {G'{f_{GG}} + R'{f_{RG}}} \right) - 2\left( {{{\rm{e}}^b} - 1} \right)({f_{GGG}}{{G'}^2} + 2R'\\\nonumber
&{f_{RGG}}G' + {G^{\prime \prime }}{f_{GG}} + {R^{\prime \prime }}{f_{RG}} + {{R'}^2}{f_{RRG}})) + 8{{\rm{e}}^{ - b}}{{R'}^2}{f_{RRR}}) + 1/\left( {\lambda  + {f_R} + 1} \right)\lambda (\frac{{{Q^2}}}{{\pi {r^4}}}\\\nonumber
& - 2{{\rm{e}}^{ - b}}{r^{ - 1}}\left( {r{{a'}^2} + \left( {2 - rb'} \right)a' - 2b' + 2r{a^{\prime \prime }}} \right){f_R}^2 - 4f + 4R{f_R} + 4G{f^{\left( {0,1} \right)}}{f_R} - \\\nonumber
&1/r8{{\rm{e}}^{ - 2b}}{f_R}(\left( {G'{f_{GG}} + R'{f_{RG}}} \right){{a'}^2} + (2({f_{GGG}}{{G'}^2} + 2R'{f_{RGG}}G' + {G^{\prime \prime }}{f_{GG}} + {R^{\prime \prime }}{f_{RG}} + \\\nonumber
&{{R'}^2}{f_{RRG}}) - 3b'\left( {G'{f_{GG}} + R'{f_R}G} \right))a' + 2{a^{\prime \prime }}(G'{f_{GG}} + RR'{f_{RG}})) - 1/r4{{\rm{e}}^{ - b}}\\\nonumber
&(2r{f_{RGG}}{{G'}^2} + (\left( {ra' - rb' + 2} \right){f_{RG}} + 4rR'{f_{RRG}})G' + 2r{G^{\prime \prime }}{f_{RG}} + ra'R'{f_{RR}} - rb'R'{f_{RR}}\\\nonumber
& + 2R'{f_{RR}} + 2r{R^{\prime \prime }}{f_{RR}} + 2r{{R'}^2}{f_{RRR}})) - 1/\left( {{r^2}\left( {\lambda  + {f_R} + 1} \right)} \right)2{{\rm{e}}^{ - 2b}}\lambda (a'({{\rm{e}}^b}R'{f_{RR}}{r^2}\\\nonumber
& + 2{{\rm{e}}^b}{f_R}^2r - 4({{\rm{e}}^b} - 3)R'{f_R}{f_{RG}} + G'\left( {{{\rm{e}}^b}{r^2}{f_{RG}} - 4\left( {{{\rm{e}}^b} - 3} \right){f_{GG}}{f_R}} \right)) + b'({{\rm{e}}^b}R'{f_{RR}}\\\nonumber
&{r^2} + 2{{\rm{e}}^b}{f_R}^2r - 4( - 3 + {{\rm{e}}^b})R'{f_R}{f_{RG}} + G'\left( {{{\rm{e}}^b}{r^2}{f_{RG}} - 4\left( {{{\rm{e}}^b} - 3} \right){f_{GG}}{f_R}} \right)) + 2( - {{\rm{e}}^b}{R^{\prime \prime }}\\\nonumber
&{f_{RR}}{r^2} - {{\rm{e}}^b}{{R'}^2}{f_{RRR}}{r^2} + 4{{\rm{e}}^b}{R^{\prime \prime }}{f_R}{f_{RG}} - 4{R^{\prime \prime }}{f_R}{f_{RG}} + {G^{\prime \prime }}(4\left( {{{\rm{e}}^b} - 1} \right){f_{GG}}{f_R} - {{\rm{e}}^b}\\\nonumber
&{r^2}{f_{RG}}) + {{G'}^2}\left( {4\left( {{{\rm{e}}^b} - 1} \right){f_{GGG}}{f_R} - {{\rm{e}}^b}{r^2}{f_{RGG}}} \right) + 4{{\rm{e}}^b}{{R'}^2}{f_R}{f_{RRG}} - 4{{R'}^2}{f_R}{f_{RRG}} + \\
&2G'R'\left( {4\left( {{{\rm{e}}^b} - 1} \right){f_R}{f_{RGG}} - {{\rm{e}}^b}{r^2}{f_{RRG}}} \right))))
\end{align}

\begin{align}\nonumber
{P_r} &= \frac{1}{{\left( {8\pi {r^4}\left( {\lambda  + {f_R} + 1} \right)\left( {2\lambda  + {f_R} + 1} \right)} \right)}}({{\rm{e}}^{ - 2b}}(4{{\rm{e}}^{2b}}\pi f\left( {\lambda  + {f_R} + 1} \right){r^4}\\\nonumber
& - 2\pi ( - 4{{\rm{e}}^b}{f_R}^3 + 4{{\rm{e}}^{2b}}{f_R}^3 - 4{{\rm{e}}^b}ra'{f_R}^3 - 4{{\rm{e}}^b}{f_R}^2 + 4{{\rm{e}}^{2b}}{f_R}^2\\\nonumber
& + {{\rm{e}}^b}{r^2}\lambda {{a'}^2}{f_R}^2 - 8{{\rm{e}}^b}\lambda {f_R}^2 + 8{{\rm{e}}^{2b}}\lambda {f_R}^2 - 4{{\rm{e}}^b}ra'{f_R}^2 - 4{{\rm{e}}^b}r\lambda a'{f_R}^2 - {{\rm{e}}^b}\\\nonumber
&{r^2}\lambda a'b'{f_R}^2 + 2{{\rm{e}}^b}{r^2}\lambda {a^{\prime \prime }}{f_R}^2 + 8{{\rm{e}}^b}a'G'{f_{GG}}{f_R}^2 - 24a'G'{f_{GG}}{f_R}^2 + 8{{\rm{e}}^b}\\\nonumber
&a'R'{f_{RG}}{f_R}^2 - 24a'R'{f_{RG}}{f_R}^2 + 4r\lambda {{a'}^2}G'{f_{GG}}{f_R} + 8{{\rm{e}}^b}a'G'{f_{GG}}{f_R} + \\\nonumber
&12{{\rm{e}}^b}\lambda a'G'{f_{GG}}{f_R} - 36\lambda a'G'{f_{GG}}{f_R} - 24a'G'{f_{GG}}{f_R} - 4{{\rm{e}}^b}\lambda b'G'{f_{GG}}\\\nonumber
&{f_R} + 12\lambda b'G'{f_{GG}}{f_R} - 12r\lambda a'b'G'{f_{GG}}{f_R} + 8r\lambda G'{a^{\prime \prime }}{f_{GG}}{f_R} + 8{{\rm{e}}^b}\lambda \\\nonumber
&{G^{\prime \prime }}{f_{GG}}{f_R} - 8\lambda {G^{\prime \prime }}{f_{GG}}{f_R} + 8r\lambda a'{G^{\prime \prime }}{f_{GG}}{f_R} + 8{{\rm{e}}^b}\lambda {{G'}^2}{f_{GGG}}{f_R} - 8\lambda \\\nonumber
&{{G'}^2}{f_{GGG}}{f_R} + 8r\lambda a'{{G'}^2}{f_{GGG}}{f_R} + 2{{\rm{e}}^{2b}}{r^2}R(\lambda  + {f_R} + 1){f_R} + 2{{\rm{e}}^{2b}}{r^2}G{f_G}\\\nonumber
&\left( {\lambda  + {f_R} + 1} \right){f_R} - 8{{\rm{e}}^b}rG'{f_{RG}}{f_R} - 2{{\rm{e}}^b}{r^2}a'G'{f_{RG}}{f_R} + 4r\lambda {{a'}^2}R'{f_{RG}}{f_R}\\\nonumber
& + 8{{\rm{e}}^b}a'R'{f_{RG}}{f_R} + 12{{\rm{e}}^b}\lambda a'R'{f_{RG}}{f_R} - 36\lambda a'R'{f_{RG}}{f_R} - 24a'R'{f_{RG}}{f_R}\\\nonumber
& - 4{{\rm{e}}^b}\lambda b'R'{f_{RG}}{f_R} + 12\lambda b'R'{f_{RG}}{f^{\left( {1,0} \right)}} - 12r\lambda a'b'R'{f_{RG}}{f_R} + 8r\lambda R'{a^{\prime \prime }}{f_{RG}}{f_R}\\\nonumber
& + 8{{\rm{e}}^b}\lambda {R^{\prime \prime }}{f_{RG}}{f_R} - 8\lambda {R^{\prime \prime }}{f_{RG}}{f_R} + 8r\lambda a'{R^{\prime \prime }}{f_{RG}}{f_R} + 16{{\rm{e}}^b}\lambda G'R'{f_{RGG}}{f_R}\\\nonumber
& - 16\lambda G'R'{f_{RGG}}{f_R} + 16r\lambda a'G'R'{f_{RGG}}{f_R} - 8{{\rm{e}}^b}rR'{f_{RR}}{f_R} - 2{{\rm{e}}^b}{r^2}a'R'{f_{RR}}\\\nonumber
&{f_R} + 8{{\rm{e}}^b}\lambda {{R'}^2}{f_{RRG}}{f_R} - 8\lambda {{R'}^2}{f_{RRG}}{f_R} + 8r\lambda a'{{R'}^2}{f_{RRG}}{f_R} - 8{{\rm{e}}^b}rG'{f_{RG}}\\\nonumber
& - 12{{\rm{e}}^b}r\lambda G'{f_{RG}} - 2{{\rm{e}}^b}{r^2}a'G'{f_{RG}} - {{\rm{e}}^b}{r^2}\lambda a'G'{f_{RG}} - {{\rm{e}}^b}{r^2}\lambda b'G'{f_{RG}} + 2{{\rm{e}}^b}{r^2}\lambda {G^{\prime \prime }}\\\nonumber
&{f_{RG}} + 2{{\rm{e}}^b}{r^2}\lambda {{G'}^2}{f_{RGG}} - 8{{\rm{e}}^b}rR'{f_{RR}} - 12{{\rm{e}}^b}r\lambda R'{f_{RR}} - 2{{\rm{e}}^b}{r^2}a'R'{f_{RR}} - {{\rm{e}}^b}{r^2}\lambda \\\nonumber
&a'R'{f_{RR}} - {{\rm{e}}^b}{r^2}\lambda b'R'{f_{RR}} + 2{{\rm{e}}^b}{r^2}\lambda {R^{\prime \prime }}{f_{RR}} + 4{{\rm{e}}^b}{r^2}\lambda G'R'{f_{RRG}} + 2{{\rm{e}}^b}{r^2}\lambda {{R'}^2}{f_{RRR}})\\
&{r^2} + {{\rm{e}}^{2b}}{Q^2}\left( {3\lambda  + {f_R} + 1} \right)))
\end{align}

\begin{align}\nonumber
{P_t} &= \frac{1}{{\left( {8\pi {r^4}\left( {\lambda  + {f_R} + 1} \right)\left( {2\lambda  + {f_R} + 1} \right)} \right)}}({{\rm{e}}^{ - 2b}}(4{{\rm{e}}^{2b}}\pi f\left( {\lambda  + {f_R} + 1} \right){r^4}\\\nonumber
& + 2\pi ({{\rm{e}}^b}{r^2}{{a'}^2}{f_R}^3 + 2{{\rm{e}}^b}ra'{f_R}^3 - 2{{\rm{e}}^b}rb'{f_R}^3 - {{\rm{e}}^b}{r^2}a'b'{f_R}^3 + 2{{\rm{e}}^b}{r^2}{a^{\prime \prime }}{f_R}^3 + {{\rm{e}}^b}{r^2}{{a'}^2}{f_R}^2\\\nonumber
& + {{\rm{e}}^b}{r^2}\lambda {{a'}^2}{f_R}^2 + 2{{\rm{e}}^b}ra'{f_R}^2 - 2{{\rm{e}}^b}rb'{f_R}^2 - 4{{\rm{e}}^b}r\lambda b'{f_R}^2 - {{\rm{e}}^b}{r^2}a'b'{f_R}^2 - {{\rm{e}}^b}{r^2}\lambda a'b'{f_R}^2\\\nonumber
& + 2{{\rm{e}}^b}{r^2}{a^{\prime \prime }}{f_R}^2 + 2{{\rm{e}}^b}{r^2}\lambda {a^{\prime \prime }}{f_R}^2 + 4r{{a'}^2}G'{f_{GG}}{f_R}^2 - 12ra'b'G'{f_{GG}}{f_R}^2 + 8rG'{a^{\prime \prime }}{f_{GG}}{f_R}^2\\\nonumber
& + 8ra'{G^{\prime \prime }}{f_{GG}}{f_R}^2 + 8ra'{{G'}^2}{f_{GGG}}{f_R}^2 + 4r{{a'}^2}R'{f_{RG}}{f_R}^2 - 12ra'b'R'{f_{RG}}{f_R}^2 + 8rR'{a^{\prime \prime }}{f_{RG}}{f_R}^2\\\nonumber
& + 8ra'{R^{\prime \prime }}{f_{RG}}{f_R}^2 + 16ra'G'R'{f_{RGG}}{f_R}^2 + 8ra'{{R'}^2}{f_{RR}}G{f_R}^2 + 4r{{a'}^2}G'{f_{GG}}{f_R} + 4r\lambda {{a'}^2}G'{f_{GG}}{f_R}\\\nonumber
& + 4{{\rm{e}}^b}\lambda a'G'{f_{GG}}{f_R} - 12\lambda a'G'{f_{GG}}{f_R} + 4{{\rm{e}}^b}\lambda b'G'{f_{GG}}{f_R} - 12\lambda b'G'{f_{GG}}{f_R} - 12ra'b'G'{f_{GG}}{f_R}\\\nonumber
& - 12r\lambda a'b'G'{f_{GG}}{f_R} + 8rG'{a^{\prime \prime }}{f_{GG}}{f_R} + 8r\lambda G'{a^{\prime \prime }}{f_{GG}}fR - 8{{\rm{e}}^b}\lambda {G^{\prime \prime }}{f_{GG}}{f_R} + 8\lambda {G^{\prime \prime }}{f_{GG}}{f_R}\\\nonumber
& + 8ra'{G^{\prime \prime }}{f_{GG}}{f_R} + 8r\lambda a'{G^{\prime \prime }}{f_{GG}}{f_R} - 8{{\rm{e}}^b}\lambda {{G'}^2}{f_{GGG}}{f_R} + 8\lambda {{G'}^2}{f_{GGG}}{f_R} + 8ra'{{G'}^2}{f_{GGG}}{f_R}\\\nonumber
& + 8r\lambda a'{{G'}^2}{f_{GGG}}{f_R} - 2{{\rm{e}}^{2b}}{r^2}R\left( {\lambda  + {f_R} + 1} \right){f_R} - 2{{\rm{e}}^{2b}}{r^2}G{f_G}\left( {\lambda  + {f_R} + 1} \right){f_R} + 4{{\rm{e}}^b}rG'{f_R}\\\nonumber
&G{f_R} + 2{{\rm{e}}^b}{r^2}a'G'{f_R}G{f_R} - 2{{\rm{e}}^b}{r^2}b'G'{f_R}G{f_R} + 4r{{a'}^2}R'{f_R}G{f_R} + 4r\lambda {{a'}^2}R'{f_R}G{f_R} + 4{{\rm{e}}^b}\lambda \\\nonumber
&a'R'{f_R}G{f_R} - 12\lambda a'R'{f_R}G{f_R} + 4{{\rm{e}}^b}\lambda b'R'{f_R}G{f_R} - 12\lambda b'R'{f_R}G{f_R} - 12ra'b'R'{f_{RG}}{f_R} - 12r\\\nonumber
&\lambda a'b'R'{f_{RG}}{f_R} + 8rR'{a^{\prime \prime }}{f_{RG}}{f_R} + 8r\lambda R'{a^{\prime \prime }}{f_{RG}}{f_R} + 4{{\rm{e}}^b}{r^2}{G^{\prime \prime }}{f_{RG}}{f_R} - 8{{\rm{e}}^b}\lambda {R^{\prime \prime }}{f_{RG}}{f_R} + 8\lambda \\\nonumber
&{R^{\prime \prime }}{f_{RG}}{f_R} + 8ra'{R^{\prime \prime }}{f_{RG}}{f_R} + 8r\lambda a'{R^{\prime \prime }}{f_{RG}}{f_R} + 4{{\rm{e}}^b}{r^2}{{G'}^2}{f_{RGG}}{f_R} - 16{{\rm{e}}^b}\lambda G'R'{f_{RGG}}{f_R} + 16\lambda \\\nonumber
&G'R'{f_{RGG}}fR + 16ra'G'R'{f_{RGG}}{f_R} + 16r\lambda a'G'R'{f_{RGG}}{f_R} + 4{{\rm{e}}^b}rR'{f_{RR}}{f_R} + 2{{\rm{e}}^b}{r^2}a'R'{f_{RR}}{f_R}\\\nonumber
& - 2{{\rm{e}}^b}{r^2}b'R'{f_{RR}}{f_R} + 4{{\rm{e}}^b}{r^2}{R^{\prime \prime }}{f_{RR}}{f_R} - 8{{\rm{e}}^b}\lambda {{R'}^2}{f_{RRG}}{f_R} + 8\lambda {{R'}^2}{f_{RRG}}{f_R} + 8ra'{{R'}^2}{f_{RRG}}{f_R}\\\nonumber
& + 8r\lambda a'{{R'}^2}{f_{RRG}}{f_R} + 8{{\rm{e}}^b}{r^2}G'R'{f_{RRG}}{f_R} + 4{{\rm{e}}^b}{r^2}{{R'}^2}{f_{RRR}}{f_R} + 4{{\rm{e}}^b}rG'{f_{RG}} + 4{{\rm{e}}^b}r\lambda G'{f_{RG}}\\\nonumber
& + 2{{\rm{e}}^b}{r^2}a'G'{f_{RG}} + {{\rm{e}}^b}{r^2}\lambda a'G'{f_{RG}} - 2{{\rm{e}}^b}{r^2}b'G'{f_{RG}} - 3{{\rm{e}}^b}{r^2}\lambda b'G'{f_{RG}} + 4{{\rm{e}}^b}{r^2}{G^{\prime \prime }}{f_{RG}}\\\nonumber
& + 6{{\rm{e}}^b}{r^2}\lambda {G^{\prime \prime }}{f_{RG}} + 4{{\rm{e}}^b}{r^2}{{G'}^2}{f_{RGG}} + 6{{\rm{e}}^b}{r^2}\lambda {{G'}^2}{f_{RGG}} + 4{{\rm{e}}^b}rR'{f_{RR}} + 4{{\rm{e}}^b}r\lambda R'{f_{RR}}\\\nonumber
& + 2{{\rm{e}}^b}{r^2}a'R'{f_{RR}} + {{\rm{e}}^b}{r^2}\lambda a'R'{f_{RR}} - 2{{\rm{e}}^b}{r^2}b'R'{f_{RR}} - 3{{\rm{e}}^b}{r^2}\lambda b'R'{f_{RR}} + 4{{\rm{e}}^b}{r^2}{R^{\prime \prime }}{f_{RR}}\\\nonumber
& + 6{{\rm{e}}^b}{r^2}\lambda {R^{\prime \prime }}{f_{RR}} + 8{{\rm{e}}^b}{r^2}G'R'{f_{RRG}} + 12{{\rm{e}}^b}{r^2}\lambda G'R'{f_{RRG}} + 4{{\rm{e}}^b}{r^2}{{R'}^2}{f_{RRR}}\\
& + 6{{\rm{e}}^b}{r^2}\lambda {{R'}^2}{f_{RRR}}){r^2} - {{\rm{e}}^{2b}}{Q^2}\left( {\lambda  + fR + 1} \right)))
\end{align}

where
\begin{equation}
{\cal R} = \frac{1}{{2{r^2}}}{{\rm{e}}^{ - b}}\left[ {4 - 4{{\rm{e}}^b} + r\left\{ {\left( {4 + ra'} \right)\left( {a' - b'} \right) + 2r{a^{\prime \prime }}} \right\}} \right]
\end{equation}
\begin{equation}
{\cal G} = \frac{{2{e^{ - 2b}}}}{{{r^2}}}\left[ {\left( {1 - {e^b}} \right){{a'}^2} + \left( {{e^b} - 3} \right)a'b' - 2\left( {{e^b} - 1} \right){a^{\prime \prime }}} \right].
\end{equation}
and
Using equation (\ref{forcharge}), we find the expression for charge, read as
\begin{align}\nonumber
Q = &\frac{{\surd [ - 1/\left\{ {{r^2}(1 + 3\lambda  + 2{\lambda ^2} + \left( {2 + 3\lambda } \right){f_R} + f_R^2)} \right\}}}{{\sqrt {\frac{{1 + 3\lambda  + {f_R}}}{{{r^4}\left( {1 + 3\lambda  + 2{\lambda ^2} + \left( {2 + 3\lambda } \right){f_R} + f_R^2} \right)}}} }}\\\nonumber
&\sqrt {2\pi } {{\rm{e}}^{ - 2A{r^2}}}( - 2{{\rm{e}}^{A{r^2}}}\left( {2{{\rm{e}}^{A{r^2}}} + A{r^2} - 3B{r^2} - 2} \right)f_R^3 + 2f_R^2(2{{\rm{e}}^{A{r^2}}} - 2{{\rm{e}}^{2A{r^2}}}\\\nonumber
& - A{{\rm{e}}^{A{r^2}}}{r^2} + 3B{{\rm{e}}^{A{r^2}}}{r^2} + 2Bg{{\rm{e}}^{2A{r^2}}}{r^2} - {{\rm{e}}^{2A{r^2}}}{r^2}RB + 4{{\rm{e}}^{A{r^2}}}\lambda  - 4{{\rm{e}}^{2A{r^2}}}\lambda  - 2A\\\nonumber
&{{\rm{e}}^{A{r^2}}}{r^2}\lambda  + 2AB{{\rm{e}}^{A{r^2}}}{r^4}\lambda  - 2{B^2}{{\rm{e}}^{A{r^2}}}{r^4}\lambda  - {{\rm{e}}^{2A{r^2}}}GB{r^2}{f_G} + 2{B^2}{f_{GGG}}G' - 2{B^2}\\\nonumber
&{{\rm{e}}^{A{r^2}}}{f_{GGG}}G' - 6ABr{f_{RG}}R' + 18{B^2}r{f_{RG}}R' + 2AB{{\rm{e}}^{A{r^2}}}r{f_{RG}}R' - 6{B^2}{{\rm{e}}^{A{r^2}}}r{f_{RG}}R' + \\\nonumber
&2{B^2}{f_{RRG}}R' - 2{B^2}{{\rm{e}}^{A{r^2}}}{f_{RRG}}R' + 4{B^2}G{f_{RG}}G'R' - 4{B^2}{{\rm{e}}^{A{r^2}}}G{f_{RG}}G'R' + 2B{f_{GG}}\\\nonumber
&\left( {\left( {A - 3B} \right)\left( { - 3 + {{\rm{e}}^{A{r^2}}}} \right)rG' - \left( { - 1 + {{\rm{e}}^{A{r^2}}}} \right){G^{\prime \prime }}} \right) + 2B{f_{RG}}{R^{\prime \prime }} - 2B{{\rm{e}}^{A{r^2}}}{f_{RG}}{R^{\prime \prime }})\\\nonumber
& + {f_R}(8Bg{{\rm{e}}^{2A{r^2}}}{r^2} + 2{{\rm{e}}^{2A{r^2}}}f{r^2} - 2{{\rm{e}}^{2A{r^2}}}{r^2}RB + 12Bg{{\rm{e}}^{2A{r^2}}}{r^2}\lambda  - 2{{\rm{e}}^{2A{r^2}}}{r^2}\\\nonumber
&RB\lambda  - 2{{\rm{e}}^{2A{r^2}}}B{r^2}\left( {1 + \lambda } \right)G{f_G} + 4{B^2}{f_{GGG}}G' - 4{B^2}{{\rm{e}}^{A{r^2}}}{f_{GGG}}G' + 16{B^2}\lambda {f_{GGG}}G'\\\nonumber
& - 16{B^2}{{\rm{e}}^{A{r^2}}}\lambda {f_{GGG}}G' - 16{B^3}{r^2}\lambda {f_{GGG}}G' + 8B{{\rm{e}}^{A{r^2}}}r{f_{RG}}G' + {B^2}{{\rm{e}}^{A{r^2}}}G{r^2}{f_{RG}}G' - AB\\\nonumber
&{{\rm{e}}^{A{r^2}}}{r^3}{f_{RG}}G' + 3{B^2}{{\rm{e}}^{A{r^2}}}{r^3}{f_{RG}}G' - 12ABr{f_{RG}}R' + 36{B^2}r{f_{RG}}R' + 4AB{{\rm{e}}^{A{r^2}}}r{f_{RG}}R' - 12{B^2}\\\nonumber
&{{\rm{e}}^{A{r^2}}}r{f_{RG}}R' - 48ABr\lambda {f_{RG}}R' + 32{B^2}r\lambda {f_{RG}}R' + 16AB{{\rm{e}}^{A{r^2}}}r\lambda {f_{RG}}R' - 16{B^2}{{\rm{e}}^{A{r^2}}}r\lambda {f_{RG}}R'\\\nonumber
& + 48A{B^2}{r^3}\lambda {f_{RG}}R' - 16{B^3}{r^3}\lambda {f_{RG}}R' + 8B{{\rm{e}}^{A{r^2}}}r{f_{RR}}R' - AB{{\rm{e}}^{A{r^2}}}{r^3}{f_{RR}}R' + 3{B^2}{{\rm{e}}^{A{r^2}}}{r^3}{f_{RR}}R'\\\nonumber
& + 4{B^2}{f_{RRG}}R' - 4{B^2}{{\rm{e}}^{A{r^2}}}{f_{RRG}}R' + 16{B^2}\lambda {f_{RRG}}R' - 16{B^2}{{\rm{e}}^{A{r^2}}}\lambda {f_{RRG}}R' - 16{B^3}{r^2}\lambda {f_{RRG}}R'\\\nonumber
& + {B^2}{{\rm{e}}^{A{r^2}}}{r^2}{f_{RRR}}R' + 8{B^2}G{f_{RG}}G'R' - 8{B^2}{{\rm{e}}^{A{r^2}}}G{f_{RG}}G'R' + 32{B^2}G\lambda {f_{RG}}G'R' - 32{B^2}{{\rm{e}}^{A{r^2}}}\\\nonumber
&G\lambda {f_{RG}}G'R' - 32{B^3}G{r^2}\lambda {f_{RG}}G'R' + 2{B^2}{{\rm{e}}^{A{r^2}}}{r^2}{f_{RRG}}G'R' + B{{\rm{e}}^{A{r^2}}}{r^2}{f_{RG}}{G^{\prime \prime }} + 4B{f_{GG}}\\\nonumber
&(r( - B( - 9 - 8\lambda  + 4B{r^2}\lambda  + {{\rm{e}}^{A{r^2}}}\left( {3 + 4\lambda } \right)) + A({{\rm{e}}^{A{r^2}}}\left( {1 + 4\lambda } \right) + 3\left( {4B{r^2}\lambda  - 1 - 4\lambda } \right)))\\\nonumber
&G' - (4B{r^2}\lambda  - 1 - 4\lambda  + {{\rm{e}}^{A{r^2}}}(1 + 4\lambda )){G^{\prime \prime }}) + 4B{f_{RG}}{R^{\prime \prime }} - 4B{{\rm{e}}^{A{r^2}}}{f_{RG}}{R^{\prime \prime }} + 16B\lambda {f_{RG}}{R^{\prime \prime }}\\\nonumber
& - 16B{{\rm{e}}^{A{r^2}}}\lambda {f_{RG}}{R^{\prime \prime }} - 16{B^2}{r^2}\lambda {f_{RG}}{R^{\prime \prime }} + B{{\rm{e}}^{A{r^2}}}{r^2}{f_{RR}}{R^{\prime \prime }}) + {{\rm{e}}^{A{r^2}}}r(r({B^2}{f_{RRR}}R' + 2({{\rm{e}}^{A{r^2}}}\\\nonumber
&\left( {1 + \lambda } \right)\left( {2Bg + f + 4Bg\lambda } \right) + {B^2}{f_{RRG}}G'R')) + B{f_{RG}}((8 - A{r^2} + Br\left( {G + 3r} \right) + \\
&12\lambda )G' + r{G^{\prime \prime }}) + B{f_{RR}}((8 - A{r^2} + 3B{r^2} + 12\lambda )R' + r{R^{\prime \prime }}))).
\end{align}
Here prime denotes the derivative with respect to radial coordinate $r$.

\vspace{0.5cm}

\bibliographystyle{unsrt}
\bibliography{mybib}
\end{document}